\documentclass[preprint,authoryear,12pt,review]{elsarticle}

\usepackage{amssymb}

\usepackage{amssymb,amsmath,amsthm}
\usepackage{graphicx,psfrag,epsf}
\usepackage{enumerate}
\usepackage{amsfonts}
\usepackage{booktabs}
\usepackage{color,soul}
\usepackage{setspace}
\usepackage{algorithm}

\newcommand{\resethlcolor}{\sethlcolor{yellow}}\resethlcolor
\definecolor{lightblue}{rgb}{.90,.95,1}
\definecolor{bluegreen}{rgb}{0,1,0.5}

\newtheorem{thm}{Theorem}[subsection]
\newtheorem{theorem}[thm]{Theorem}

\newtheorem{corollary}[thm]{Corollary}

\def \m {\mathfrak{m}}

\def \A {\mathcal{A}}
\def \R {\mathbb{R}}
\def \Z {\mathbb{Z}}
\def \Diam {\operatorname{Diam}}
\def \Var {\operatorname{Var}}
\def \Cov {\operatorname{Cov}}

\journal{Journal of Econometrics}

\begin{document}

\begin{frontmatter}



\title{On spherical Monte Carlo simulations for  multivariate normal probabilities}


\author[ncu]{Huei-Wen Teng}
\author[nctu]{Ming-Hsuan Kang}
\author[ncu]{Cheng-Der Fuh}

\address[ncu]{Graduate Institute of Statistics, National Central University,
Jhongli City,Taiwan.}

\address[nctu]{Department of Applied Mathematics, National Chiao Tung University,
Hsinchu,Taiwan.}

\begin{abstract}

The calculation of multivariate normal probabilities is of great importance in many statistical and economic applications. This paper proposes a spherical Monte Carlo method with both theoretical analysis and numerical simulation. First, 
the multivariate normal probability is rewritten via an inner radial integral and an outer spherical integral by the spherical transformation. 
For the outer spherical integral, we apply an integration rule by randomly rotating a predetermined set of well-located points. To find the desired set, we derive an upper bound for the variance of the Monte Carlo estimator and propose a set which is related to the kissing number problem in sphere packings. For the inner radial integral, we employ the idea of antithetic variates and identify certain conditions so that variance reduction is guaranteed. 
Extensive Monte Carlo experiments on some probabilities calculation confirm these claims.

\end{abstract}

\begin{keyword}
spherical \sep simulation \sep variance reduction \sep kissing number \sep lattice.
\MSC C

\end{keyword}

\end{frontmatter}


\section{Introduction}
\label{sec:intro}

Efficient and precise calculation for the multivariate normal probability is of critical importance
in many disciplines. To name a few, the multinomial probit model used in econometrics and biometrics has cell probabilities that are negative orthant probabilities. In financial industry implementation
of the CreditMetrics model, the joint migration probabilities in credit migration
model are rectangle probabilities for bivariate normal distributions, cf. \cite{JPMorgan1997}. The estimation of Value-at-Risk for risk management considered in \cite{GHS2000} requires calculation of multivariate probability of an ellipsoid.
In multiple comparisons, multivariate normal probabilities are also considered in \cite{Hsu1996}. For more examples and applications, the reader is referred to \cite{GB2009} for a recent summary of multivariate normal distribution and multivariate $t$ distribution.

Motivated by these applications, we investigate efficient calculation for the multivariate
normal probability. That is, for an indicator function  $I_{\A}(x)$  with a support set $\A$  in $\R^d$, we seek more efficient computation of the following probability
\begin{eqnarray}\label{normal}
p_{\A}=P\{X \in \A \}= \int_{\R^d} I_{\A}(x) \frac{1}{\sqrt{(2 \pi)^d|\Sigma|}} e^{-\frac{1}{2} (x-\mu)' \Sigma^{-1}
(x-\mu)} dx,~~~{\rm for}~\	\A \subset {\R}^d,
\end{eqnarray}
where $'$ denotes the transpose, $X=(X_1,\cdots,X_d)'$ is a $d$-dimensional non-singular multivariate normal distribution with mean
vector $\mu$ and covariance matrix $\Sigma$, denoted by $X\sim N_d(\mu,\Sigma)$, and $|\Sigma|$ denotes the determinant of $\Sigma$.

Standard approaches of calculating (\ref{normal}) include classical analytic approximation, numerical integration and Monte Carlo method. Instead of using analytic approximation and
numerical method, cf. \cite{Miwaetal2003} and \cite{Craig2008}, which are usually more suitable for low dimensional cases, in this paper, we study Monte Carlo method.
Although Monte Carlo method is easy to implement and can overcome the curse of dimensionality, its convergence rate is rather slow (proportional to $1/\sqrt{d}$).
Therefore, additional variance reduction methods are required.  Typical methods for variance reduction include antithetic variates, Latin hypercube sampling, and primitive Monte Carlo method,
cf. \cite{Genz1992}, \cite{Genz1993}, \cite{HMR1996}, \cite{V1997}, \cite{GB2002}, and among others.
A comparison study of alternative sampling methods can be found in \cite{SA2004}.
A survey on existing methods is in \cite{GB2009}.

Monte Carlo methods using spherical transformation have been studied in the literature. For instance, \cite{Deak1980} and \cite{Deak2000} used this transformation as the basis for calculating
multivariate normal probabilities, and \cite{FW1994} proposed a transformation on the unit hypercube to generate points uniformly distributed on the sphere.
\cite{MonahanGenz1997} proposed a Monte Carlo simulation method for Bayesian computation, to which
the authors used randomized extended simplex design for the spherical integral and Simpson weights for the radial integral.

In this paper, we propose a spherical Monte Carlo method with both theoretical analysis and numerical simulations.
There are two aspects in this study. First, because the spherical integral requires generating unit vectors uniformly on the sphere, one way to
improve the Monte Carlo efficiency is to use a randomly rotated predetermined set of unit vectors. For this purpose, we give a criterion to select such a set on the unit sphere,
which involves the minimal distance of any two points in the set and the cardinality of the set.
Among all sets with the same minimal distance, the desired optimal set is the one with maximal cardinality.
Especially, when the minimal distance equals one, finding the desired optimal set is linked to the kissing number problem in sphere packings.

Next, for the radial integral, we offer a variance reduction technique employing the idea of antithetic variates.
For this purpose, we introduce the idea of {\it central symmetry} for the set of basis points and {\it central antisymmetry} for the event of simulation.
To the best of our knowledge, this seems to be a first step to provide sufficient conditions for antithetic variates on spheres.
To illustrate the proposed method, simulation studies are given for orthants, rectangles, and ellipsoids probabilities for multivariate normal distributions. The simulation results confirm these claims.

The rest of this paper is organized as follows. In \S \ref{sec:ourmethod} we propose the spherical Monte Carlo method with antithetic variates.
\S \ref{sec:spherepacking} links the proposed set of basis points to a sphere packing problem and related spherical $t$-designs, and discusses practical implementation for high dimensional cases.
Simulation results are given in \S \ref{sec:numericalresults}. \S \ref{sec:conclusion} concludes. The proofs are deferred to Appendix.


\section{The proposed spherical Monte Carlo method}
\label{sec:ourmethod}

For easy presentation, we split this section as three subsections. \S \ref{sub:pf} formulates the problem, \S \ref{sub:si}
presents the spherical integral, and \S \ref{subsection:estimators}
studies the radial integral.

\subsection{Problem formulation}\label{sub:pf}
Let $I_\A:\R^d\rightarrow \R$ be an indicator function with a support set $\A$. Denote the probability density function (pdf) of a multivariate normal random variable $N_d(\mu,\Sigma)$ as
$$\phi(x;\mu,\Sigma) = \frac{1}{\sqrt{(2\pi)^d|\Sigma|}}e^{-(x-\mu)' \Sigma^{-1}(x-\mu)},$$
where $\mu$ is the mean vector and $\Sigma$ is the variance-covariance matrix.
The desired multivariate normal probability of the region $\A$ is an integral of the form in (\ref{normal}).

Note that a $d$-variate normal random variable $X\sim N_d(\mu,\Sigma)$ can be expressed by
$X = \mu + \Gamma Z$, where $Z$ is a $d$-variate standard normal random variable, and $\Gamma$ is the lower triangle matrix such that
$\Sigma= \Gamma \Gamma'$, the so-called Cholesky decomposition of $\Sigma$. By change of variables, $x=\mu + \Gamma z$, we rewrite (\ref{normal}) as
\begin{equation}
\label{eq:p2}
p_\A =  \int_{\R^d} I_{\tilde{\A}}(z) \phi(z) dz,
\end{equation}
where $\phi(\cdot)$ is the standard normal density and $\tilde{\A}=\Gamma^{-1}(\A-\mu)$.
Abuse the notation a little bit, we denote $\tilde{\A}$ by $\A$ in the rest of the paper.

By using the spherical transformation, a point $z\in\R^d$ can be written as
$(r,u)$, where $r$ is the radius and $u$ is the unit vector of $z$. Then, $I_{{\A}}(z)= I_{{\A}}(r, {u})$ and (\ref{eq:p2}) equals
\begin{equation}
\label{eq:p3}
p_\A =\int_{S^{d-1}} \int_{0}^{\infty} I_{{\A}}(r, {u}) \left( \frac{1}{ \sqrt{(2\pi)^d}} e^{-r^2/2} r^{d-1} \right) dr dA = \frac{1}{\text{Area}(S^{d-1})}\int_{S^{d-1}} f_{\A}({u}) du,
\end{equation}
where $dA$ denotes the differential area on the unit sphere $S^{d-1}$, and the inner radial integral is
\begin{equation}
\label{eq:f(u)}
f_{\A}({u})= \text{Area}(S^{d-1}) \int_{0}^{\infty} I_{{\A}}(r, u) \left( \frac{1}{ \sqrt{(2\pi)^d}} e^{-r^2/2} r^{d-1} \right) dr = \int_0^\infty I_{\A}(r,u)k_d(r)dr.
\end{equation}
Here, $\text{Area}(S^{d-1})={2 \sqrt{\pi^d}}/{\Gamma(\frac{d}{2})}$ is the surface area of $S^{d-1}$,  and $k_d(\cdot)$ is the pdf of a $\chi$-distribution with degrees of freedom $d$, denoted by $\chi(d)$.
A random unit vector $u$ in $\R^d$ is a vector uniformly distributed on $S^{d-1}$, denoted by $u \sim U(S^{d-1})$.
Therefore, generating a sample from a standard normal distribution $z$ is equivalent to independently generating a radius $r\sim \chi(d)$ and a unit vector $u\sim U(S^{d-1})$, and setting $z=ru$.

To rewrite (\ref{eq:p2}) using the spherical transformation, an alternative approach is to consider the spherical integral as the innermost integral as in \cite{MonahanGenz1997}.
Here, we take the radial integral as the innermost integral because the radial integral is of one dimension and its calculation is simple.
For a simple region $\A$ like rectangles, orthants and ellipsoids, when the unit vector $u$ is fixed,
the inner radial integral has a closed-form formula in terms of cumulative distribution function (cdf) of a $\chi$-distribution, cf.
\cite{Deak2000}. For general regions, the inner radial integral can be calculated via Monte Carlo methods
or numerical methods, cf. \cite{DavisRabinowitz1984}. Numerical quadrature methods produce biased estimators in general and are out of the scope of this paper. Here, we only focus on Monte Carlo methods.

The outer spherical integral using Monte Carlo method requires generating a random unit vector $u$. A straightforward method to generate $u$ is to generate $d$
independent standard normal random variables to have a vector in $\R^d$ and then normalize the vector by its length.
A more efficient algorithm can be found in \cite{FW1994} by generating just $(d-1)$ random numbers to get $u$.

The crude spherical Monte Carlo estimator is
\begin{equation}
\label{eq:hatp}
\hat{p} = I_{{\A}}(r,u),
\end{equation}
where $r\sim \chi(d)$, $u \sim U(S^{d-1})$, and $r$ and $u$ are independent. The crude spherical Monte Carlo estimator with antithetic variates is
\begin{equation}
\label{eq:hatp_AT}
\hat{p}_{AT} = \frac{I_{{\A}}(r,u)+I_{{\A}}(r,-u)}{2}.
\end{equation}

\subsection{Variance reduction on the spherical integral}\label{sub:si}

One way to obtain an efficient spherical Monte Carlo estimator for the spherical integral is taking the average value of a randomly rotated predetermined finite set of unit vectors, denoted by $V$.
More precisely, let $O(d)$ be the orthogonal group of $\R^d$ consisting of all $d\times d$ matrices $D$ satisfying that $DD'$ equals the identity matrix.
There are a unique probability measure on $O(d)$ which is invariant under left multiplication by all elements of $O(d)$ and
a random orthogonal matrix $T$ which is uniformly distributed on $O(d)$ with respect to this probability measure, denoted by $T\sim U(O(d))$.
Then randomly rotating a set $V$ simply means to apply a random orthogonal matrix $T$ to every element of $V$.

The standard algorithm to generate a random orthogonal matrix can be described as follows, cf. \cite{He}.
First, generate a random $d\times d$ matrix whose entries are independently standard normal random variables.
Then, apply Gram-Schmidt method to the column vectors and obtain the desired random orthogonal matrix.

One important feature of random orthogonal matrix is that if we fix a unit vector and let a random orthogonal matrix act on it,
the resulting vector is uniformly distributed on $S^{d-1}$. In other words, for a continuous function $h(u)$ on $S^{d-1}$,  and any unit vector $v \in S^{d-1}$, we have 
\def \Area {\operatorname{Area}}
\begin{equation} \label{formula:1}
\frac{1}{\Area(S^{d-1})}\int_{S^{d-1}} h(u) du =  \int_{O(d)} h(Tv) dT.
\end{equation}
Here $dT$ is the unique left-invariant probability measure on $O(d)$ mentioned above.

More efficient algorithms using only $(d-1)(d+2)/2$ standard normal random variables can be found in \cite{St}, \cite{DiaconisShahshahani1987}, and \cite{AOU}.

A spherical Monte Carlo estimator using $V$ is
\begin{equation}
\label{eq:hatp_SR}
\hat{p}^V = \frac{1}{|V|}\sum_{v\in V}I_{{\A}}(r_v,Tv),
\end{equation}
where $r_v$ are independent random variables with $\chi(d)$ distribution, and $T \sim U(O(d))$ which is independent of $r_v$.
When the innermost radial integral can be calculated explicitly, i.e., $f_{\A}(u)$ can be expressed in terms of the cdf of $\chi(d)$, an estimator using $V$ is
\begin{equation}
\label{eq:hatp_S}
\hat{p}_*^V = \frac{1}{|V|}\sum_{v\in V} f_{\A}(Tv),
\end{equation}
where $T\sim U(O(d))$. By using the fact of conditioning, indeed, $f_{\A}(T v)$ equals $E[I_{{\A}}(r_v,T v)|T]$, and thus enjoys smaller variance compared with $\hat{p}^V$.

To have efficient simulation, the crucial step involves the selection of $V$ to minimize the variance of $\hat{p}^V$.
For this purpose, we propose a criterion of determining the desired set $V$ in the following, and provide a feasible solution that related
to the maximal kissing number problem in sphere packings in \S 3. 

To begin with, for a finite subset $V$ of the unit sphere $S^{d-1}$, to characterize the variance of the estimator for the spherical integral,
we need to define some notations as follows. First, denote $d(\cdot,\cdot)$ as the usual Euclidean distance function on $\R^d$, and let
$$ d_{\min}(V)= \min\{ d(v,v') \,|\, v,v' \in V, v\neq v'\}$$
be the minimal distance of any two points in $V$. For a region $\A$ in $S^{d-1}$, define the diameter of $\A$ as
$$\Diam(\A) = \sup \{d(x,x') \,|\, x,x' \in \A\}.$$
It is easy to see that when $\Diam(\A)<  d_{\min}(V) $, the intersection of $V$ and $A$ is either the empty set or a single point.
\def \m {\mathfrak{m}}

Let $\pi$ be the normalized measure on $S^{d-1}$ induced from the Lebesgue measure on $\R^d$ so that $\pi(S^{d-1})=1$. Note that $\pi$ is a probability measure and it is invariant under the action of $O(d)$.
For a region $\A$ in $S^{d-1}$, define an estimator on $O(d)$ as follows,
\begin{equation}
\label{union}
g^V_{\A}(T)  =  \sum_{v\in V} \frac{1}{|V|} I_{\A} (Tv),
\end{equation}where $T\sim U(O(d))$.
Suppose $\Diam(\A)<  d_{\min}(V)$. Let $TV$ denote the set consisting of $Tv$ for all $v\in V$. Since orthogonal matrices preserve the Euclidean distance function, we have $d_{\min}(TV)=d_{\min}(V)$ and
$|{\A} \cap TV | = 1 \mbox{ or } 0.$ On the other hand, the estimator $g^V_{\A}(T)$ is unbiased. Therefore, we have
$$g^V_{\A}(T) = \left\{\begin{array}{ll}\frac{1}{|V|},&\mbox{with probability } |V|\pi({\A}),\\ 0,&\mbox{otherwise.}\end{array}\right.$$
Note that $\Diam(\A)<  d_{\min}(V)$ implies
$ \pi({\cal A}) < 1/|V|$. Moreover,
\begin{eqnarray}\label{var}
\Var(g^V_{\A}(T)) = \left(\frac{1}{|V|}\right)^2 |V|\pi(\A) - \pi^2(\A) = \frac{\pi(\A)}{|V|}-\pi^2(\A).
\end{eqnarray}

If  $\Diam(\A) \geq  d_{\min}(V)$,
we can decompose $\A$ and obtain the following

\begin{theorem}
\label{thm:upperbound}
Let $\A$ be a region of $S^{d-1}$ which can be written as a disjoint union of the subregions $\A_1,\cdots,\A_{N}$ so that $\Diam(\A_i)<  d_{\min}(V)$ for all $i=1,\cdots,N$.
Then the variance of $g^V_{\A}(T)$  satisfies
$$ \Var(g^V_{\A}(T)) \leq \pi(\A)-\pi^2(\A) + \left(\frac{N}{|V|}-1\right)\pi(\A).$$
\end{theorem}

The proof of Theorem \ref{thm:upperbound} will be given in Appendix A. 

When $N$ is the minimal number such that the decomposition condition in Theorem \ref{thm:upperbound} holds,
denote the upper bound in  Theorem \ref{thm:upperbound} by $C(V,\A)$ for short.
Note that $C(V,\A)$ reduces to (\ref{var}), the exact variance, when $N=1$ in Theorem \ref{thm:upperbound}. In general case,
if $N < |V|$, $C(V, \A)$ is less than $\pi(\A)-\pi^2(\A)$,
the variance of the crude Monte Carlo estimator $\hat{p}$.
Next we immediately have

\begin{corollary}
\label{cor:1}
For given two finite subsets $V$ and $V'$ of $S^{d-1}$, $C(V,\A)\leq C(V',\A)$ for any region $\A$ in $S^{d-1}$ if one of the following two conditions holds:
\begin{enumerate}
\item $|V|=|V'|$ and $d_{\min}(V)>d_{\min}(V')$;
\item $d_{\min}(V)=d_{\min}(V')$ and $|V|>|V'|$.
\end{enumerate}
\end{corollary}

Fix a given region $\A$, the upper bound $C(V,\A)$ depends only on the minimal distance between any two points in $V$, and the cardinality of the set $V$.
Corollary \ref{cor:1} suggests two approaches to minimize the upper bound of the variance of the estimator $\hat{p}^V$ defined in (\ref{eq:hatp_SR}):
\begin{enumerate}
\item[] Case 1: among all sets $V$ of the same cardinality, the set with  maximal $d_{\min}(V)$ minimizes $C(V,\A)$;
\item[] Case 2: among all sets $V$ of the same $d_{\min}(V)$, the set with maximal cardinality minimizes $C(V,\A)$.
\end{enumerate}

Although there is no general method to construct such set $V$ in either case, useful results are available. Other than the construction of $V$ based
on Case 1\footnote{In Case 1, for $d$ from three to five and $|V|<150$, sets and known maximal $d_{\min}(V)$ are reported in the Sloane website http://www2.research.att.com/~njas/packings/},
here we provide a solution based on Case 2. Note that in the case of $d_{\min}(V)=2$, $V$ must have exactly two points, since the distance between any two unit vectors is less than or equal to two,
and the equality holds only when they form an antipodal pair. This shows that the method of antithetic variates is optimal under the above criterion for $d_{\min}(V)=2$.
For the case of $d_{\min}(V)=1$, constructing a set with maximal cardinality is related to {\it the kissing number problem in sphere packings}.
Details of this linkage will be discussed in \S \ref{kissing}.

\subsection{Variance reduction on the radial integral} \label{subsection:estimators}

The efficiency of simulators may be improved, for a given number of Monte
Carlo draws, by using of antithetic variates. A straightforward application of antithetic variates on sphere can be found in \cite{Deak1980}. Here we have a different approach by applying the idea of antithetic variates for a set of points on sphere $S^{d-1}$.
For this purpose, we first introduce the concepts of central symmetry and central antisymmetry. A set $V$ on $S^{d-1}$
is called {\it centrally symmetric} if for all $v$ in $V$, $-v$ also lies in $V$. In this case, we can decompose $V$ as the disjoint union of $V^+$ and $-V^+$.
Here  $V^+$ consists of all vectors in $V$ whose first non-zero coordinates is positive. It is easy to see that the property of centrally symmetric
is preserved under the action of $O(d)$.
Motivated by the idea of antithetic variates, a proposed spherical Monte Carlo estimator using a centrally symmetric set $V$ with antithetic variates is
\begin{eqnarray}\label{eq:hatp_SR,AT}
\hat{p}^V_{AT} = \frac{1}{|V|} \sum_{v \in V^+} I_{{\A}}(r_{v},Tv )+I_{{\A}}(r_{v},-T v),
\end{eqnarray}
where $r_v\sim \chi(d)$ for $v\in V$, $T\sim U(O(d))$, and $r_v$'s and $T$ are independent.
We remark that the proposed set $V$ generated by the shortest nonzero vectors of a lattice as will be described in \S 3 is centrally symmetric so that $\hat{p}^V_{AT}$ in (\ref{eq:hatp_SR,AT}) is well defined.

\def \Cov {\operatorname{Cov}}

It is known that certain conditions are required in Cartesian coordinate to insure antithetic variates enjoy lower variance, cf. \cite{Ross2006}. Likewise,
an additional condition is needed in the spherical case.
A region $\A$ is called {\it centrally antisymmetric} if $\A$ and $(-\A)$ are disjoint. It is easy to see that the property of centrally antisymmetric
is also preserved under the action of $O(d)$.
As a result, we have the following

\begin{theorem}
\label{thm:at}
If the set $V$ on $S^{d-1}$ is centrally symmetric and the event $\A$ is centrally antisymmetric, then
$\Var(\hat{p}^V_{AT}) \leq \Var( \hat{p}^V )$.
\end{theorem}

The proof of Theorem \ref{thm:at} will be given in Appendix B. 

Theorem \ref{thm:at} shows that the variance of $\hat{p}^V_{AT}$ can be reduced when $\A$ is centrally antisymmetric.
As will be shown in \S \ref{subsec:results}, for certain central antisymmetric sets, although $\Var(\hat{p}^V_{AT}) =\Var(\hat{p}^V)$ as a mathematical fact,
$\hat{p}^V_{AT}$ is preferred because it requires less number of independently drawn radii and thus enjoys lower computational cost compared with $\hat{p}^V$.
For demonstration purposes, \S \ref{subsec:results} provides an explicit central antisymmetric set, for which $\hat{p}^V_{AT}$ enjoys smaller variance than $\hat{p}^V$.

\section{Practical implementation}

\label{sec:spherepacking}
This section first explores the link of an optimal $V$ based on Case 2 in Corollary \ref{cor:1} to the kissing number problem in sphere packings, and then makes a connection to spherical $t$-designs. \S \ref{subsec:highd} discusses possible approaches for high-dimensional cases.

\subsection{The linkage to the kissing number problem in sphere packings}
\label{kissing}
\begin{figure}
\begin{center}
\includegraphics[height= 5 cm]{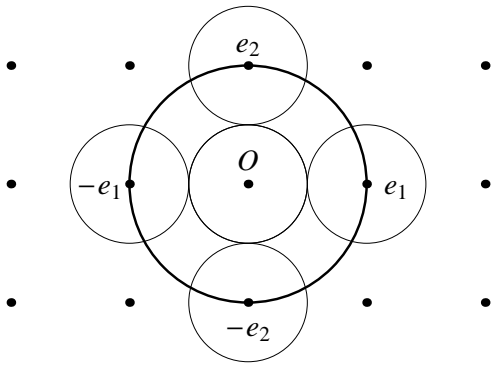}
\end{center}
\caption{Lattice $\Z^2$ and the corresponding sphere arrangement with kissing number four. $V_{\Z^2} =\{\pm e_1, \pm e_2\}$.}
\label{fig:Z2}
\end{figure}

\begin{figure}
\begin{center}
\includegraphics[height= 5cm]{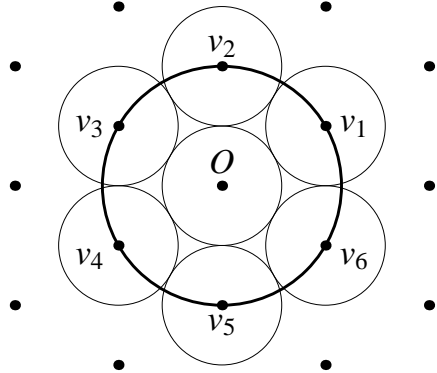}
\end{center}
\caption{Lattice $A_2$ and the corresponding sphere arrangement with kissing number six. $V_{A_2} =\{v_1,\ldots, v_6\}$.}
\label{fig:A2}
\end{figure}

For a sphere packing, a kissing number is defined as the number of non-overlapping spheres that can be arranged such that they each touch another given sphere. Here all spheres must have the same radius.
The kissing number problem seeks the maximal possible kissing number in a sphere packing.
Figures \ref{fig:Z2} and \ref{fig:A2} demonstrate two sphere arrangements in $\R^2$ with a kissing number equal to four and six, respectively. Moreover, the later one has the maximal kissing number.

Now for a finite subset $V$ of $S^{d-1}$ with $d_{\min}(V)=1$,
we can construct an associated sphere packing from $V$ as follows. Consider all spheres centred at $v\in V$ with radius $1/2$
and one sphere centred at the original with radius $1/2$.  It is easy to see that the number $|V|$ equals the kissing number of this sphere packing, and, as a result,
to find a $V$ with maximal $|V|$ with $d_{\min}(V)=1$ is exactly the same as finding a sphere packing producing the maximal kissing number.

A next question is regarding how to obtain a sphere arrangement with maximal kissing number. This question is quite difficult in geometry, but has been solved partially
so that we can directly utilize existing results. Most of known maximal kissing numbers in sphere packings come from the study of lattices.
Given a basis of $\R^d$, a {\it lattice} $L$ is defined as the set of all integral linear combinations of the basis.
Let $d_{\min}$ be the shortest distance among all pairs of points in $L$. Then one can obtain a sphere arrangement
by setting spheres with radius $\frac{1}{2}d_{\min}$ centred at all points of $L$. In this case, every sphere kisses
the same number of spheres. Especially, for the sphere centred at the origin, the number of spheres kissing it equals
the number of shortest non-zero vectors in $L$. For short, we simply call the shortest non-zero vectors by shortest vectors in the rest of the paper.  To obtain $V$ from a lattice $L$, we simply
collect all the shortest vectors and normalize them to unit vectors, and denote it by $V_L$.

For example, the integer lattice $\Z^d$ is generated by the standard basis $\{e_1,\cdots,e_d\}$ and consists of all integer points.
In this case, the kissing number of $\Z^d$ equals $2d$, and the set of shortest vectors 
$$V_{\Z^d}=\{ \pm e_1,\cdots,\pm e_d\}.$$
Figure \ref{fig:Z2} shows the lattice and the derived  sphere arrangement for $d=2$.

Another example considers the lattice $A_2$, which is generated by the basis $v_1=(\frac{\sqrt{3}}{2},\frac{1}{2})$ and $v_2=(0,1)$.
The lattice and the associated sphere arrangement are depicted in Figure \ref{fig:A2}. Note that the kissing number of $A_2$ is six,
which is indeed the maximal kissing number for $d=2$. Moreover, we obtain
$$V_{A_2}=\left\{ \left(\pm \frac{\sqrt{3}}{2},\pm \frac{1}{2} \right),\left(0,\pm 1 \right)\right\}.$$

It is difficult to prove if a kissing number is maximal for an arbitrary $d$, cf. \cite{ConwaySloane1999}.
For $d$ from two to eight, lattices that produce the known maximal kissing number are the lattices $A_2$, $A_3$, $D_4$, $D_5$, $E_6$, $E_7$, and $E_8$.
These lattices are called root lattices which come from semisimple Lie algebras and these symbols stands for the name of the corresponding Lie algebra.
Explicit construction of these lattices can be found in \cite{ConwaySloane1999}. We can thus form the desired $V$ from these lattices, by taking the normalized shortest vectors in these lattices.
Table \ref{tab:cardinality} lists the cardinality of each $V_L$ from the corresponding lattices for $d$ from two to eight, 16 and 24.

It is worth mentioning that the set $V_L$ above also forms
a spherical $t$-design for some $t>0$. That is
$$ \frac{1}{\text{Area}(S^{d-1})}\int_{S^{d-1}} g(u) du =  \frac{1}{|V_L|}\sum_{v \in V_L} g(v)$$
for all real-valued continuous functions $g(u)$ on the sphere which are restrictions of polynomial functions of degree less than or equal to $t$ in $\R^d$.
This means if the function $g(u)$ is a polynomial function of degree less than or equal to $t$, its integral over the sphere $S^{d-1}$ is simply an average of function values of these points. To some extent, elements in $V_L$ behave as a quadrature point that allows
the integral of polynomials equals to a finite sum of polynomial values. On the other hand, if the function $g(u)$ can be
approximated by a polynomial well, than its integral can be approximated as an average of these functional values. Then choosing $V_L$ as the sample set provides a good approximation of the desired integral.
In other words, the spherical Monte Carlo estimator
$g^{V_L}(T):= \sum_{v \in V_L} \frac{1}{|V_L|} g(Tv)$ produces zero variance for estimating a polynomial function of degree less than or equal to $t$ in $\R^d$, under the multivariate normal probability density and hence enjoys a small variance if $g(\cdot)$
can be well approximated by a polynomial of degree less than or equal to $t$ in $\R^d$.
Table \ref{tab:cardinality} lists the associated $t$ for the proposed $V_L$ as a spherical $t$-design.

\subsection{High-dimensional cases}

\label{subsec:highd}

In principle, our proposed method is feasible for higher dimensions once a suitable $V$ is selected. Theorem
\ref{thm:upperbound} suggests the use of $V$ with the maximal cardinality given $d_{\min}(V)=1$, which is related to the maximal kissing number problem in sphere packings, and can be constructed via a lattice in some cases. The lattice with the maximal presently known kissing number, denoted by $L_d^{\max}$, for dimension up to 40 can be found on the website http://www.math.rwth-aachen.de/$\sim$Gabriele.Nebe/LATTICES/.
For demonstration purposes, we report results for $d$ from two to eight in \S \ref{subsec:results}.

Nevertheless, some difficulties arise to construct $V$ from the lattice $L^{\max}_d$: not all of the optimal lattices are well-studied lattices so that  their shortest vectors can not be constructed generically. Moreover,  the size of $V$ may be too large for practical implementation in high-dimensional cases. For example, the maximal kissing number is 196,560 for $d=24$, which seems be formidable for practical implementation. 

For high-dimensional cases, a simple (but not optimal based on our criterion in Theorem \ref{thm:upperbound}) approach is to use a certain family of lattices. For example, we consider three lattices $\Z_d$, $D_d$, and $A_d$ in this paper, which are defined as follows. 
\begin{itemize}
\item $\Z_d$ is generated by the standard basis $e_1,\cdots,e_d$ of $\R^d$ and $V_{\Z_d}=\{\pm e_1,\cdots, \pm e_d\}.$
\item $D_d$ is generated by $(e_1-e_2), (e_2-e_3),\cdots,(e_{d-1}-e_d)$ and $(e_{d-1}+e_d).$ $V_{D_d}$ consists of all permutations of $\frac{1}{\sqrt{2}}(\pm 1, \pm 1 , 0 ,\cdots, 0)$ and there are $2d(d-1)$ such vectors.
\item Identify $\R^d$ as the hyperplane $x_1+ \cdots +x_{d+1}=0$ in $\R^{d+1}$. Then $A_d$ is generated by $(e_1-e_2), (e_2-e_3),\cdots,(e_{d}-e_{d+1})$ and $V_{A_d}$ consists of all permutations of $\frac{1}{\sqrt{2}}( 1, - 1 , 0 ,\cdots, 0) \in \R^{d+1}$ and there are $d(d+1)$ such vectors.
\end{itemize}
Therefore, spherical packings generating by lattices $\Z_d$, $A_d$ and $D_d$ have kissing numbers of $2d$, $d(d+1)$ and $2d(d-1)$, respectively. 
An additional advantage of using these lattices is that the associated $V$ of these three families of lattices can be constructed generically. However, as will be compared in \S \ref{subsec:Deak}, spherical estimators employing $V$ based on these lattices produce larger variances. As a remark, the spherical estimator $\hat{p}^{V_{D_d}}_*$ in our framework equals to the orthonomalized-2 estimator in \cite{Deak2000}, although this connection was not revealed in the original paper.


\begin{table}
\caption{The lattice $L_d^{\max}$, the cardinality of its normalized shortest vectors $V$ , and the associated $t$ of $V$ as a spherical $t$-design, for various dimensions $d$}
\label{tab:cardinality}
\centering
\begin{tabular}{cccccccccc}
\toprule
$d$  & 2 & 3 & 4 & 5 & 6 & 7 & 8 & 16 & 24 \\
\midrule
$L_d^{\max}$ & $A_2$ &  $A_3$ & $D_4$ & $D_5$ & $E_6$ & $E_7$ & $E_8$ & Barnes-Wall lattice & Leech lattice  \\
$|V|$ & 6 & 12 & 24 & 40 & 72 & 126 & 240& 4320 & 196560 \\
$t$ & 5 & 3 & 5 & 3 & 5 & 5 & 7 & 7 & 11\\
\bottomrule
\end{tabular}

\end{table}

\section{Simulation studies}\label{sec:numericalresults}
For easy presentation,
\S \ref{subsec:algorithm} presents algorithms for each estimator, and discusses the computational cost and penalized variance ratios.
A simulation design is given in \S \ref{subsec:plan}, and the associated numerical results are summarized in \S \ref{subsec:results}. Finally, \S \ref{subsec:Deak}
compares the performance of $V$ 
arising from the proposed lattice $L_d^{\max}$ with various lattices for $d=16$ and 24.

\subsection{Algorithms and the computational cost}
\label{subsec:algorithm}

To fairly compare the efficiency of the proposed estimators, we report variance ratios and penalized variance ratios as measures in our simulation studies.
The variance ratio is defined as the variance of a crude Monte Carlo estimator divided by that of an estimator of interest. An estimator with variance ratio larger than one is hence more efficient than the crude Monte Carlo estimator.
Since the quantity of interest in this paper is the multivariate normal probability and the integrand consists of an indicator function, we restrict the computational cost to be the possible smallest number of independent random numbers required to generate one realization for the estimator.
By incorporating the computational cost, we define the penalized variance of an estimator as the product of its variance and associated computational cost, and define the penalized variance ratio of an estimator as the penalized variance of the crude Monte Carlo estimator divided by that of the estimator of interest.

To estimate $p_\A$ given in (\ref{normal}), we present explicit algorithms for each estimator described in this paper.
Let $M$ be the Monte Carlo sample size.
Because the crude spherical Monte Carlo estimator is indeed the same as the usual crude Monte Carlo estimator, a procedure for generating $\hat{p}$ is given as follows.
\begin{enumerate}
\item Generate $x^{(i)}\sim N_d(\mu,\Sigma)$ for $i=1,\;\ldots,\;M$.
\item Set $\hat{p} = \frac{1}{M}\sum_{i=1}^MI_\A(x^{(i)})$.
\end{enumerate}
Similarly, a procedure for generating $\hat{p}_{AT}$ in (\ref{eq:hatp_AT}) is given as follows.
\begin{enumerate}
\item Generate $z^{(i)}\sim N_d(0, I)$ for $i = 1, \ldots, M$.
\item Set $x^{+(i)}=\mu + \Gamma z^{(i)}$ and $x^{-(i)}=\mu-\Gamma z^{(i)}$.
\item Set $\hat{p}_{AT} = \frac{1}{2M} \sum_{i=1}^MI_{\A}(x^{+(i)})+I_{\A}(x^{-(i)})$.
\end{enumerate}
It is clear that the computational cost of $\hat{p}$ and $\hat{p}_{AT}$ are both equal to $d$.

For a given centrally symmetric subset $V=\{v_1,\cdots,v_{|V|}\}$ of $S^{d-1}$,
we have estimators $\hat{p}^V$, $\hat{p}^V_{AT}$, and $\hat{p}^V_*$.
An algorithm for calculating $\hat{p}^V$ is implemented as follows.
\begin{enumerate}
\item Generate a random orthogonal matrix, $T^{(i)}\sim U(O(d))$, for $i=1,\;\ldots,\;M$.
\item Generate a radius, $r^{(i)}_j\sim \chi(d)$, for $i=1,\ldots,M$, and $j=1,\ldots, |V|$.
\item Set $z^{(i)}_j = r^{(i)}_j T^{(i)}v_j$ for $i=1,\ldots,M$, and $j=1,\ldots, |V|$.
\item Set $x^{(i)}_j = \mu + \Gamma z^{(i)}_j$ for $i=1,\ldots,M$, and $j=1,\ldots, |V|$.
\item Set $\hat{p}^V = \frac{1}{M|V|} \sum_{i=1}^M\sum_{j=1}^{|V|} I_\A(x^{(i)}_j).$
\end{enumerate}
Because a realization of $\hat{p}^V$ requires to generate one random orthogonal matrix and $|V|$ independent radii, the computational cost of $\hat{p}^V$ is $(d+2)(d-1)/2+|V|$.
Recall in \S \ref{sub:si} that the computational cost to generate a $d\times d$ orthogonal matrix is $(d+2)(d-1)/2$.

Note that the the cardinality of the proposed $V$ is even since it is centrally symmetric. Let $V^+$ consists of all vectors in $V$ whose first non-zero coordinate is positive,
and $V$ is decomposed as the disjoint union of $V^+$ and $-V^+$.
Without loss of generality, we assume that $V^+=\{v_1,\cdots,v_{|V|/2} \}$.
An algorithm for calculating $\hat{p}^V_{AT}$ is outlined as follows.
\begin{enumerate}
\item Generate a random orthogonal matrix, $T^{(i)}\sim U(O(d))$, for $i=1,\;\ldots,\;M$.
\item Generate a radius, $r^{(i)}_j\sim \chi(d)$, for $i=1,\ldots,M$, and $j=1,\ldots, |V|/2$.
\item Set $z^{+(i)}_j = r^{(i)}_j T^{(i)}v_j$ and set $z^{-(i)}_j = -z^{+(i)}_j$ for $i=1,\ldots,M$, and $j=1,\ldots, |V|/2$.
\item Set $x^{+(i)}_j = \mu + \Gamma z^{+(i)}_j$ and $x^{-(i)}_j = \mu + \Gamma z^{-(i)}_j$ for $i=1,\ldots,M$, and $j=1,\ldots, |V|/2$.
\item Set $\hat{p}^V_{AT} = \frac{1}{2M|V|} \sum_{i=1}^M\sum_{j=1}^{|V|/2} I_\A(x^{+(i)}_j)+I_\A(x^{-(i)}_j).$
\end{enumerate}
Because generating a realization of $\hat{p}^V_{AT}$ requires a generation of one random orthogonal matrix and $|V|/2$ radii, the computational cost of $\hat{p}^V_{AT}$ is $(d+2)(d-1)/2+|V|/2$.

For calculating $\hat{p}^V_*$, recall that \S \ref{sec:ourmethod} has shown that
$
\int_{\R^d} I_\A(x)\phi(x;\mu,\Sigma)dx = \int_{O(d)} f_{\tilde{\A}}(u) dT,
$ with
$f_{\tilde{\A}}(u) =\int_0^{\infty} I_{\tilde{\A}}(r,u)d(r)dr$ and $\tilde{\A} = \Gamma^{-1}(A-\mu)$.
When explicit formulas for ${f}_{\tilde{\A}}(u)$ exists (in terms of cdf of chi distributions), $\hat{p}^V_*$ can be calculated as follows.
\begin{enumerate}
\item Generate a random orthogonal matrix, $T^{(i)}\sim U(O(d))$, for $i=1,\;\ldots,\;M$.
\item Set $u^{(i)}_j=T^{(i)}v_j$ for $i=1,\ldots,M$, and $j=1,\ldots, |V|$.
\item Set $\hat{p}^V_* = \frac{1}{M|V|} \sum_{i=1}^M\sum_{j=1}^{|V|}f_{\tilde{A}}(u^{(i)}_j)$.
\end{enumerate}
Note that the calculation of $f_{\tilde{A}}(u)$ involves the cdf of a $\chi$-distribution, and such additional computational effort can not be reflected in the definition of the penalized variance ratios. In our simulation studies in \S \ref{sec:numericalresults}, we do not report the penalized variance ratio for $\hat{p}^V_*$. Table \ref{tab:cost} summarizes the computational cost for four estimators: $\hat{p}$, $\hat{p}_{AT}$, $\hat{p}^V$, and $\hat{p}^V_{AT}$.

\begin{table}
\caption{The computational cost of various estimators and dimensions $d$ with the proposed $V=V_{L^{\max}_d}$ in \S \ref{sec:spherepacking}.}~\\
\label{tab:cost}

\centering
\begin{tabular}{lrrrrrrrc}\toprule
$d$ & 2	 & 	3	 & 	4	 & 	5	 & 	6	 & 	7	 & 	8	 &  $d$	\\\midrule
$\hat{p}$  & 2	 & 	3	 & 	4	 & 	5	 & 	6	 & 	7	 & 	8 	 & $d$	\\
$\hat{p}_{AT}$  & 2	 & 	3	 & 	4	 & 	5	 & 	6	 & 	7	 & 	8	 &	$d$ \\
$\hat{p}^V$ & 8	 & 	17	 &  33 & 54 & 92	 & 	153	 & 	275	 & $\frac{1}{2}(d+2)(d-1)+|V_L|$		\\
$\hat{p}^V_{AT}$  & 5	 & 	11	 & 	21	 & 	34	 & 	56	 & 	90	 & 	155	 &	$\frac{1}{2}(d+2)(d-1)+\frac{1}{2}|V_L|$\\
\bottomrule
\end{tabular}
\end{table}

\subsection{Simulation design}
\label{subsec:plan}
In this subsection, we would like to compare the efficiency of various estimators in calculating $P\{X\in \mathcal{A}\}$ for some set $\mathcal{A}$ and $X\sim N_d(0, \Sigma)$.

Similar to the simulation settings in \cite{V1997}, we consider three types of covariance models: the identity covariance matrix, the one-factor model, and the AR(1) model.
For the one-factor model with parameter $\rho$, the covariant matrix is set as $\Sigma=(\rho_{ij})$ with $\rho_{ij}=1$ if $i=j$ and $\rho_{ij}=\rho$ if $i\neq j$.
For the AR(1) model with parameter $\rho$, the covariant matrix is set as $\Sigma=(\rho_{ij})$ with $\rho_{ij}=\rho^{|i-j|}$ for all $i$ and $j$.
For the one-factor model and the AR(1) model, four different values of $\rho$, $-0.1$, $0.1$, $0.2$, and $0.3$, are used. All these $\rho$'s produce positive definite covariance matrices.

We consider three types of regions: ellipsoids, orthants and rectangles.
For each type of the region, we denote three sets as follows,
\begin{itemize}
\item ellipsoid regions:
$E_1= \{ x \in \R^d, (x-b)'(x-b)\leq 1, b=(1,0,\ldots,0)'\}$, and $E_2= \{ x \in \R^d, (x-b)'(x-b)\leq 1, b=(0.5,0,\ldots,0)'\}$, $E_3= \{ x \in \R^d, (x-b)'(x-b)\leq 1, b=(1,1,\ldots,1)'\}$;
\item orthant regions: $O_1=[-\infty, 0]^d$, $O_2=[-\infty,1]^d$,
and $O_3=[-\infty,-1]^d$;
\item rectangular regions: $R_1=[-1,1]^d$, $R_2=[0,2]^d$, and $R_3=[0.5,1.5]^d$.
\end{itemize}

Note that the interiors of $R_2$, $O_1$, $O_3$, and $E_2$ are centrally antisymmetric. It is clear that the property of centrally antisymmetric
is preserved under changing coordinate via a linear transformation. Therefore when the region $\A$ is centrally antisymmetric, $\tilde{\A}=\Gamma^{-1}(\A)$ remains centrally antisymmetric.
Also, we run the simulation for $d$ from two to eight. Overall, there are seven dimensions, five estimators, nine covariance structures,
and nine sets of interest, and a total of 2,835 combinations. The Monte Carlo sample size is $10,000$ for each case.

\subsection{Numerical results}
\label{subsec:results}

Table \ref{tab:vr} summarizes averaged (penalized) variance ratios among all types of the covariance structures as described in \S 4.2. Overall, the efficiency of $\hat{p}^V_*$ is the highest, followed by $\hat{p}^V$ and $\hat{p}^V_{AT}$, and $\hat{p}_{AT}$.
On average, $\hat{p}_{AT}$ produces variance ratio about two for all dimensions, whereas $\hat{p}^V$,  $\hat{p}^V_{AT}$, $\hat{p}^V_*$ produce higher variance ratios for higher dimensions. In addition, the variance ratio of $\hat{p}^V_*$ is dramatically larger in all cases.
This numerical evidence suggests that if the innermost radial integral $f_{\A}(u)$ can be calculated explicitly, $\hat{p}^V_*$ are preferred.

On the other hand, although calculating $\hat{p}^V_\ast$ does not need generating the radius, its calculation requires
the calculation of the cdf of the $\chi$-distribution, which may be computationally demanding.
Thus, we omit listing averaged penalized variance ratios for $\hat{p}^V_*$. Again, the averaged penalized variance ratios for $\hat{p}_{AT}$ are about two for all dimensions, but increase slightly for $\hat{p}^V_{AT}$ and $\hat{p}_{AT}$ for higher dimensions.
Although averaged variance ratios of $\hat{p}^V_{AT}$ are slightly smaller than those of $\hat{p}^V$ in most cases,
averaged penalized variance ratios of $\hat{p}^V_{AT}$ are larger. 
In other words, these numerical results suggest that although $\hat{p}^V_{AT}$ do not yield higher variance ratio than $\hat{p}^V$, accounting for the expense of drawing independent random variables, $\hat{p}^V_{AT}$ are preferred.

Our simulation studies show that $\hat{p}^V_*$ work particularly well for elliptical regions. In fact, when $\A$ is a circle centred at the original, $f_{\A}(u)$ is about the same for all $u$, and thus $\hat{p}^V_*$ produces a dramatically high variance ratio. (When the covariance matrix is the identity matrix, $f_{\A}(u)$ is identical for all $u$, and is a zero-variance estimator.) This exemplifies a feature that a spherical Monte Carlo estimator, $\hat{p}^V_*$, is very efficient for certain ellipsoid regions (regardless of the selection of $V$).

To investigate how the property of centrally antisymmetric affect the results, we average the (penalized) variance ratios among all centrally antisymmetric
sets ($R_2$, $O_1$, $O_3$, and $E_2$). See Table \ref{tab:anti} for details. In this case, the variance ratios of $\hat{p}^V_{AT}$ are quite close to those of $\hat{p}^V$.
This provides a numerical evidence that the property of centrally antisymmetric does matter when incorporating the idea of antithetic variates.

\begin{table}
\caption{Averaged variances ratios (VR) and averaged penalized variance ratios (PVR) using various estimators for calculating $P\{X\in \mathcal{A}\}$ over all covariance structures for three types of region: ellipsoid region (E), orthant region (O), and rectangular region (R), at various dimensions $d$. Monte Carlo sample size is 10,000. }
\label{tab:vr}

\centering
\begin{tabular}{lrrrrrrr}\toprule
	&	\multicolumn{4}{c}{VR}						&		\multicolumn{3}{c}{PVR}\\
	\cmidrule(rl){2-5}   					\cmidrule(rl){6-8}
	&	$\hat{p}_{AT}$	&	$\hat{p}^V$	&	$\hat{p}^V_{AT}$	&	$\hat{p}^V_{*}$	&	$\hat{p}_{AT}$	&	$\hat{p}^V$	&	$\hat{p}^V_{AT}$	\\\midrule
\multicolumn{8}{l}{$d=2$}\\										
E	&	2.76 	&	11.32 	&	8.38 	&	3152940.68 	&	2.76 	&	2.83 	&	3.35\\
O	&	2.60 	&	16.42 	&	15.96 	&	743.63 	&	2.60 	&	4.10 	&	6.38 	 \\
R	&	2.03 	&	12.75 	&	11.70 	&	702.34 	&	2.03 	&	3.19 	&	4.68 	 \\
\multicolumn{8}{l}{$d=3$}\\										
E	&	2.56 	&	22.92 	&	17.59 	&	21295621.93 	&	2.56 	&	4.04 	&	4.80\\
O	&	2.32 	&	24.90 	&	23.43 	&	209.71 	&	2.32 	&	4.39 	&	6.39 	 \\
R	&	1.79 	&	20.57 	&	18.65 	&	384.36 	&	1.79 	&	3.63 	&	5.09 	 \\
\multicolumn{8}{l}{$d=4$}\\
E	&	2.36 	&	41.47 	&	34.53 	&	41843229.01 	&	2.36 	&	5.03 	&	6.58\\
O	&	2.22 	&	47.65 	&	44.00 	&	302.56 	&	2.22 	&	5.78 	&	8.38 	 \\
R	&	1.72 	&	37.13 	&	33.15 	&	340.00 	&	1.72 	&	4.50 	&	6.31 	 \\
\multicolumn{8}{l}{$d=5$}\\
E	&   2.21 	&	69.83 	&	59.82 	&	6570613.74 	&	2.21 	&	6.47 	&	8.80 	 \\
O	&	2.19 	&	71.73 	&	64.96 	&	414.07 	&	2.19 	&	6.64 	&	9.55 	 \\
R	&	1.69 	&	54.70 	&	48.03 	&	404.46 	&	1.69 	&	5.06 	&	7.06 	 \\
\multicolumn{8}{l}{$d=6$}\\
E	&	2.14 	&	118.96 	&	108.45 	&	40738761.06 	&	2.14 	&	7.76 	&	11.62\\
O	&	2.16 	&	123.19 	&	111.29 	&	736.32 	&	2.16 	&	8.03 	&	11.92 	 \\
R	&	1.68 	&	92.48 	&	80.51 	&	742.81 	&	1.68 	&	6.03 	&	8.63 	 \\
\multicolumn{8}{l}{$d=7$}\\
E	&	2.07 	&	203.63 	&	190.59 	&	376655202.42 & 2.07 &	9.32 &	14.82\\
O	&	2.16 	&	206.96 	&	187.88 	&	1487.83 	&	2.16 	&	9.47 	&	14.61 	 \\
R	&	1.65 	&	153.15 	&	133.24 	&	1296.48 	&	1.65 	&	7.01 	&	10.36 	 \\
\multicolumn{8}{l}{$d=8$}\\
E	&	2.03 	&	387.05 	&	362.58 	&	19598639060.99 	&	2.03 	&11.26 	&	18.71 \\
O	&	2.14 	&	398.90 	&	361.15 	&	4638.70 	&	2.14 	&	11.60 	&	18.64 	 \\
R	&	1.61 	&	284.54 	&	245.32 	&	2750.45 	&	1.61 	&	8.28 	&	12.66 	 \\\bottomrule
\end{tabular}
\end{table}

\begin{table}
\caption{Averaged variance ratios (VR) and averaged 
penalized variance ratios (PVR) using estimators 
$\hat{p}^V$ and $\hat{p}^V_{AT}$ for calculating $P\{X\in \mathcal{A}\}$ 
over all centrally antisymmetric sets ($R_2$, $O_1$, $O_3$, and $E_2$) and covariance structures at  various dimensions $d$. Monte Carlo sample size is 10,000. }
\label{tab:anti}
\centering
\begin{tabular}{lrrrr}\toprule
& \multicolumn{2}{c}{VR} & \multicolumn{2}{c}{PVR} \\
\cmidrule(rl){2-3}   \cmidrule(rl){4-5}
$d$	&	$\hat{p}^V$	&	$\hat{p}^V_{AT}$	&	$\hat{p}^V$	&	$\hat{p}^V_{AT}$	 \\\midrule
2	&	18.58 	&	17.79 	&	4.65 	&	7.12 	 \\
3	&	27.12 	&	26.03 	&	4.79 	&	7.10 	 \\
4	&	48.76 	&	47.08 	&	5.91 	&	8.97 	 \\
5	&	68.84 	&	66.94 	&	6.37 	&	9.84 	 \\
6	&	112.92 	&	110.67 	&	7.36 	&	11.86 	 \\
7	&	179.77 	&	178.26 	&	8.23 	&	13.86 	 \\
8	&	328.28 	&	323.12 	&	9.55 	&	16.68 	 \\\bottomrule
\end{tabular}
\end{table}

Note that due to an interesting feature in these centrally antisymmetric sets $R_2$, $O_1$, $O_3$, and $E_2$,
the averaged (penalized) variance ratios in Table \ref{tab:anti} for $\hat{p}^V$ and $\hat{p}^V_{AT}$ are indeed the same. To explain why,
let $\A$ currently be one of these four sets. Given a $v \in V$, if $r v $ belongs in $\A$, $-r'v $ does not belong in $\A$
for all $r'$. As a result, a constituent of the spherical estimator $\hat{p}^V_{AT}$, $I_{\A}(r_v, T_v)+I_\A(r_v, -Tv)$ and 
 a constituent of the spherical estimator $\hat{p}^V$,
$I_{\A}(r_{v1}, T_v)+I_\A(r_{v2}, -Tv)$, both become zero or one.
This leads to that both estimators have the same variance. Therefore, the inequality in Theorem \ref{thm:at} is indeed an equality for these four sets.

To provide a reasonable set showing that $\hat{p}^V_{AT}$ enjoys a lower variance, consider a set $S=[-1,-1/2]\times [-1,1]^{d-1} \cup[0,1/2]\times [-1,1]^{d-1}$.
Clearly, this set is centrally antisymmetric, and, in addition, it allows that when using antithetic variate, a constituent of estimator $\hat{p}^V_{AT}$, $I_{\A}(r_v, T_v)+I_\A(r_v, -Tv)$, would take values of zero and one.
On the contrary, when no antithetic variate is used, a constituent of the estimator $\hat{p}^V$, $I_{\A}(r_{v1}, T_v)+I_\A(r_{v2}, -Tv)$, would take values of zero, one, and two.
Because these two estimators are unbiased, the latter would have a larger variance. Table \ref{tab:anti2} demonstrates the desired results by showing that the averaged
variance ratios of $\hat{p}^V_{AT}$ are larger than those of $\hat{p}^V$. Although the improvement is of a moderate scale, accounting for the computational cost,
$\hat{p}^V_{AT}$ is preferred in terms of the penalized variance ratios.

\begin{table}
\caption{Averaged variance ratios (VR) and averaged 
penalized variance ratios (PVR) for calculating $P\{X\in S\}$ 
with $S=[-1,-1/2]\times [-1,1]^{d-1} \cup[0,1/2]\times [-1,1]^{d-1} $ 
over all covariance structures using estimators 
$\hat{p}^V$ and $\hat{p}^V_{AT}$ at various dimensions $d$. 
Monte Carlo sample size is 10,000. }~\\
\label{tab:anti2}
\centering
\begin{tabular}{lrrrr}\toprule
& \multicolumn{2}{c}{VR} & \multicolumn{2}{c}{PVR} \\
\cmidrule(rl){2-3}   \cmidrule(rl){4-5}
$d$	&	$\hat{p}^V$	&	$\hat{p}^V_{AT}$	&	$\hat{p}^V$	&	$\hat{p}^V_{AT}$	 \\\midrule2	&	6.77 	&	8.93 	&	1.69 	&	3.57 	 \\
3	&	13.64 	&	15.32 	&	2.41 	&	4.18 	 \\
4	&	26.74 	&	28.45 	&	3.24 	&	5.42 	 \\
5	&	43.42 	&	44.74 	&	4.02 	&	6.58 	 \\
6	&	76.16 	&	79.16 	&	4.97 	&	8.48 	 \\
7	&	132.16 	&	134.28 	&	6.05 	&	10.44 	 \\
8	&	249.22 	&	252.68 	&	7.25 	&	13.04 	 \\\bottomrule
\end{tabular}
\end{table}

\subsection{Comparison of $\hat{p}^V_*$ with $V$ generated by various lattices}
\label{subsec:Deak}

Table \ref{tab:HD} lists the cardinality and variance ratio of spherical estimators using $V$ arising from various lattices for calculating $P(Z\in O_2)$, with $O_2=[-\infty, 1]^d$ as defined in \S \ref{subsec:plan}. For simplicity, we just focus on the spherical estimator $\hat{p}^V_*$, because it enjoys lower variances than other spherical estimators due to conditioning. To compromise the effect of cardinality for a given $V$, the variance ratio penalized by the cardinality is also reported. The proposed lattices $L_d^{\max}$ based on Theorem \ref{thm:upperbound} are the Barnes-Wall lattice for $d=16$ and the Leech lattice for $d=24$. The explicit construction of the shortest vectors of these lattices can be also founded in \cite{ConwaySloane1999}.

It is clear that the cardinality of the proposed $V_{L_d^{\max}}$ is dramatically larger than the other three counterparts. The variance reduction by using $V_{L_d}^{\max}$ is substantial. As for the variance ratio penalized by the cardinality, the spherical Monte Carlo estimator employing the propsoed $V_{L_d^{\max}}$ remains the most competitive one at a factor of about three to four.

\begin{table}
\caption{The cardinality, variance ratio, and variance ratio
penalized by cardinality, in calculating $P\{Z\in O_2\}$ using estimator $\hat{p}^V_*$ with $V$ arising from various lattices
for dimensions $d=16$ and $24$. Monte Carlo sample size is 100.}
\label{tab:HD}
\centering
\begin{tabular}{lrrrr}\toprule
$d\backslash$ Lattice	 &	$\Z_d$	 &	$A_d$	 &	$D_d$	 &	$L_d^{\max}$	 \\
\midrule
\multicolumn{5}{l}{cardinality}									 \\
16	 &	32	 &	272	 &	480	 &	4320	 \\
24	 &	48	 &	600	 &	1104	 &	196560	 \\
\multicolumn{5}{l}{VR}									 \\
16	 &	116.3 	 &	840.9 	 &	1248.3 	 &	40650.8 	 \\
24	 &	146.9 	 &	1296.7 	 &	1972.0 	 &	1535820.0 	 \\
\multicolumn{5}{l}{VR-cardinality}	 		 		 		 		 \\
16	 &	3.6 	 &	3.1 	 &	2.6 	 &	9.4 	 \\
24	 &	3.1 	 &	2.2 	 &	1.8 	 &	7.8 	 \\
\bottomrule
\end{tabular}

\end{table}

\section{Conclusion}
\label{sec:conclusion}

Clearly, the proposed spherical methods exemplify the omnibus method for calculating multivariate normal
probabilities that we seek.  To explain the superior performance of the proposed integration methods, we look at the differences in integrating the radial and spherical parts.
To reduce the variance for the spherical integral,, instead of sampling one point randomly, we randomly rotate a set of points
on the sphere which is constructed from a sphere packing with the maximal kissing number.
Moreover, we employ the idea of antithetic variates and show that it can have further variance reduction under certain conditions.
Simulation results confirm the theoretical results that our method always provides substantial variance reduction in some cases.

Although the proposed spherical methods give superior efficiency,
this is only part of the story in the sense that we only
study the case of multivariate normal probabilities.
For more general cases such as the calculation of multivariate $t$ distribution, Dirichlet distributions, elliptical copulas, and even the high-dimensional integrals with general function,
we need to utilize the proposed spherical method with certain features of the distributions,
and the best implementation in this paper can be adapted to various circumstances.
After some exploration, the choice of spherical integral rule,
the interval size of the radial integral, and perhaps the radial integral rule itself can be adjusted. Of course, the choice of $V$ depends
on the development of kissing numbers in sphere packings, an alternative is to employ basis points chosen based on Case 1.
Other possibility is to consider unequal weights for the basis points, which is related to weighted spherical $t$-designs.
Using a general method such as sampling from a $\chi$-distribution for the radial part allows for any reasonable sort of tail
behaviour, while it remains competitive with other
approaches when the tail of the distribution is not normal.

In conclusion, being able to evaluate multivariate normal probabilities of
higher dimensions at low cost opens up new avenues of research on limited
dependent variables in time series analysis, panel data, and spatial models.
This one-step estimator becomes more efficient
when more adjacent periods are included in the construction of the step, at the
computational cost of evaluating higher-order integral.
In this paper, we assume that the underlying random variables are independent over time. A more
challenging project is to model the time dependence used in, for
example, financial industry implementation of the CreditMetrics model and the computation of Basel III incremental
risk charge (IRC). Since our method highlights the two aforementioned key features of spherical and radial integrations in general settings, it is also expected that this method can be applied to option pricing, Greeks letters calculation, and correlated
default probabilities evaluation among others.

\appendix

\section{Proof of Theorem \ref{thm:upperbound}}
\label{pr:thm1}

The proof of Theorem \ref{thm:upperbound} depends on the calculation of variance in {\rm (\ref{union})}.
To this end, we first consider the covariance of any two random variables appeared in {\rm (\ref{union})}.
Let $\A$ and $\A'$ be two disjoint regions of $S^{d-1}$ satisfying that $\Diam(\A)$ and $\Diam(\A')$ are both less than $d_{\min}(V)$. Because the covariances of $g^V_{\A}(T)$ and $g^V_{\A'}(T)$
is difficult to calculate, we would like to find an upper bound of the covariance instead. Since $g^V_{\A}(T)$ and $g^V_{\A'}(T)$ are simple step functions, we have
\begin{eqnarray*}
&~& E[g^V_{\A}(T) g^V_{\A'}(T)] = E\bigg( \sum_{v \in V} \frac{1}{|V|} I_{\A} (Tv) \sum_{v' \in V} \frac{1}{|V|} I_{\A'} (Tv') \bigg) \\
&=& \frac{1}{|V|^2} \sum_{v \in V} \sum_{v' \in V} E (  I_{\A} (Tv) I_{\A'} (Tv') )
= \frac{1}{|V|^2} \sum_{v, v'\in V}  P\{ Tv \in {\A}, Tv' \in {\A'} \}.
\end{eqnarray*}
Since all points $v$ in $V$ are predetermined and $Tv$ is simply a rotation on the sphere,
therefore
\[ \sum_{v, v'\in V} P\{ Tv \in {\A}, Tv' \in {\A'} \} \leq
 \sum_{v, v'\in V} \min( P\{ Tv \in \A\}, P\{ Tv' \in \A' \}).
\]
Recall that when $v$ is fixed, $Tv$ is uniformly distributed on $S^{d-1}$
which implies that
$$  P\{ Tv \in \A\}=\pi(\A) \quad \mbox{and} \qquad P\{ Tv' \in \A' \} = \pi(\A')$$
and
\begin{eqnarray*}
E[g^V_{\A}(T) g^V_{\A'}(T)]
\leq \frac{1}{|V|^2} \min(|V| \pi(\A), |V| \pi(\A')) \leq \frac{1}{2|V|}(\pi(\A)+\pi(\A')).
\end{eqnarray*}
Hence,
$$\Cov(g^V_{\A}(T), g^V_{\A'}(T)) \leq \frac{1}{2|V|}(\pi(\A)+\pi(\A'))-\pi(\A)\pi(\A').$$
Then, we have
\begin{eqnarray*}
\Var(g^V_{\A}(T)) &=& \Var(\sum_{j=1}^N g^V_{\A_j})
= \sum_{j=1}^N \Var(g^V_{\A_j}) + \sum_{i\neq j} \Cov(g^V_{\A_i}, g^V_{\A_j})\\
&\leq& \sum_{j=1}^N \frac{\pi(\A_j)}{|V|} -\pi^2(\A_i) + \sum_{i\neq j}\left(\frac{1}{2|V|}(\pi(\A_i)+\pi(\A_j))-\pi(\A_i)\pi(\A_j)\right)\\
&=& \frac{1}{|V|}\pi(\A)-\pi^2(\A) +\frac{1}{2|V|}\sum_{i\neq j}(\pi(\A_i)+\pi(\A_j))\\
&=&  \frac{1}{|V|}\pi(\A)-\pi^2(\A) +\frac{1}{2|V|}\sum_{j=1}^{N}2(N-1)\pi(\A_i)\\
&=& \frac{1}{|V|}\pi(\A)-\pi^2(\A)+\frac{N-1}{|V|}\pi(\A)\\
&=& \pi(\A)-\pi^2(\A) + \left(\frac{N}{|V|}-1\right)\pi(\A).
\end{eqnarray*}~\\

\section{Proof of Theorem \ref{thm:at}}
\label{pr:at}

To have the proof, we first note that
\begin{align}\label{varcov}
 &|V| \left(\Var(\hat{p}^V_{AT}) - \Var( \hat{p}^V) \right)\\
=&\sum_{v \in V^+} \Cov(I_{{\A}}(r_{v},Tv ),I_{{\A}}(r_{v},-T v)) - \Cov(I_{{\A}}(r_v,Tv ),I_{{\A}}(r_{-v},-T v)). \nonumber
\end{align}

If $\A$ is centrally antisymmetric, then for any $\chi(d)$-distributed independent random variables $r_1, r_2$, and $\chi(d)$-distributed random variable $r$, we have
\begin{align*}
& \Cov\left( I_{{\A}}(r_1,u ), I_{{\A}}(r_2,-u ) \right)- \Cov \left( I_{{\A}}(r,u ), I_{{\A}}(r,u ) \right)\\
=& E[I_{{\A}}(r_1,u ) I_{{\A}}(r_2,-u )] - E[I_{{\A}}(r,u )I_{{\A}}(r,-u )]\\
=&   E[I_{{\A}}(r_1,u) I_{{\A}}(r_2,-u )] \geq 0,
\end{align*}
which implies that
\begin{equation} \label{condition:cov}
 \Cov \left( I_{{\A}}(r,u ), I_{{\A}}(r,u ) \right) \leq \Cov\left( I_{{\A}}(r_1,u ), I_{{\A}}(r_2,-u ) \right).
\end{equation}

Followed by (\ref{formula:1}), we can rewrite (\ref{varcov}) as
\begin{eqnarray*}
&~& \sum_{v \in V^+} \Cov(I_{{\A}}({r_{v}},u ),I_{{\A}}({r_{v}},-u)) - \Cov(I_{{\A}}({r_{v}},u ),I_{{\A}}({r_{-v}},-u))\\
&=& \frac{|V|}{2} \left( \Cov\left( I_{{\A}}(r,u ), I_{{\A}}(r,-u ) \right)- \Cov \left( I_{{\A}}(r_1,u ), I_{{\A}}(r_2,u ) \right) \right)
\end{eqnarray*}
which is less than or equal to zero by (\ref{condition:cov}).

\clearpage

\bibliographystyle{elsarticle-harv}
\bibliography{ref}







\end{document}